\documentclass[aps,prb,twocolumn,superscriptaddress,floatfix]{revtex4-2}

\usepackage[T1]{fontenc}
\usepackage[utf8]{inputenc}

\usepackage{graphicx}
\usepackage{amsmath,amssymb,mathtools,amsfonts,bm}
\usepackage{xcolor} 
\usepackage{braket}
\usepackage{siunitx}

\usepackage{hyperref} 

\DeclareSIUnit{\kcal}{\kilo\cal}

\begin{document}

\title{Revealing Short- and Long-range Li-ion diffusion in Li$_2$MnO$_3$ from finite-temperature dynamical mean field theory}

\author{Alex Taekyung Lee}
\affiliation{Department of Chemical engineering, University of Illinois at Chicago, Chicago, IL 60608, USA}
\affiliation{Materials Science Division, Argonne National laboratory, Lemont, IL 60439, USA}

\author{Kristin A. Persson} 
\affiliation{Environmental Energy Technologies Division, Lawrence Berkeley National Laboratory, Berkeley, California 94720, United States}
\affiliation{Department of Materials Science and Engineering, University of California Berkeley, Berkeley, California 94704, United States}

\author{Anh T. Ngo} 
\affiliation{Department of Chemical engineering, University of Illinois at Chicago, Chicago, IL 60608, USA}
\affiliation{Materials Science Division, Argonne National laboratory, Lemont, IL 60439, USA}

\date{\today}

\begin{abstract}
Li$_2$MnO$_3$ is a key component of Li-excess layered cathodes of the form
$(1-x),\mathrm{LiMO_2} + x,\mathrm{Li_2MnO_3}$ ($M$ = Mn, Ni, Co, \dots), yet its role in setting
Li-ion transport limitations remains under debate. Here we combine DFT+$U$, finite-temperature
DFT+DMFT with a continuous-time quantum Monte Carlo impurity solver, and nudged-elastic-band (NEB)
calculations to study Li$^{+}$ migration in paramagnetic Li$_2$MnO$_3$ in the presence of a single Li
vacancy. Evaluating DMFT total energies along the DFT+$U$ NEB geometries reveals that dynamical
correlations strongly renormalize the lowest-barrier processes, reducing the activation energies to
$E_a = 0.18$ eV for the shortest-range hop and $E_a = 0.50$ eV for the next-lowest
(transport-controlling) step. The 0.18 eV barrier quantitatively reproduces the short-range activation
energy from $\mu^{+}$SR, while the 0.50 eV barrier is consistent with the long-range transport scale
extracted from ac-impedance measurements. This single-vacancy, paramagnetic DMFT description thus
provides a unified interpretation of local and macroscopic probes without invoking clustered vacancy
configurations or strong extrinsic disorder, consistent with nearly stoichiometric Li$_2$MnO$_3$
powders. More broadly, our results highlight finite-temperature dynamical correlations as an essential
ingredient for predicting ionic migration energetics in correlated oxide electrodes.

\end{abstract}

\maketitle

\section{Introduction}

Rising energy- and power-density targets for lithium-ion batteries have intensified the need for cathode materials that combine high capacity with strong rate capability \cite{GoodenoughCM2010,GoodenoughJACS2013,ManthiramACSCS2017,ManthiramNatcomm2020}. Within Li-excess layered oxides, Li$_2$MnO$_3$ is a key building block in composite cathode chemistries of the form $(1-x),\mathrm{LiMO_2} + x,\mathrm{Li_2MnO_3}$ ($M$ = Mn, Ni, Co, \dots) and is widely used to access high operating voltages ($\sim$4.4--5.0~V) at low cost \cite{Yabuuchi2011,SeoNchem2016,YeMT2014,Liu2016IJES}. Understanding the mechanisms and limitations of Li-ion mobility in Li$_2$MnO$_3$ is therefore crucial for interpreting rate limitations in Li-rich layered cathodes and for guiding the design of improved electrode architectures.

Experimental estimates of Li-ion migration in Li$_2$MnO$_3$ exhibit a pronounced dependence on the probe and the associated length scale \cite{Nakamura20101359,Sugiyama2013,Yu2012RSCAdv}. At the microscopic (local) scale, $\mu^+$SR measurements on nearly stoichiometric Li$_2$MnO$_3$ powders over 2--500~K report a low activation energy, $E_a = 0.156$~eV, for short-range Li jumps in the high-temperature paramagnetic regime ($T \gg T_N$) \cite{Sugiyama2013}. In contrast, ac-impedance measurements on Li$_2$MnO$_3$ ceramic powders yield a substantially larger apparent barrier, $E_a \approx 0.46$~eV, for long-range Li-ion transport \cite{Nakamura20101359}. Taken together, these results suggest a hierarchy of activation energies for local versus percolating Li motion even in nominally pristine Li$_2$MnO$_3$.

First-principles nudged-elastic-band (NEB) studies based on density functional theory
(DFT)+$U$ have provided valuable insight into the topology of Li-ion migration pathways
in Li$_2$MnO$_3$ \cite{ShinCM2016,KongJMCA2015,WangJMCA2017,Sarkar2017}.
The DFT+$U$+NEB study by Shin \emph{et al.} showed that, for Li$^+$ migrating in the
presence of a single Li vacancy, the calculated barriers are 0.5--0.8 eV for intralayer hops
and 0.64--0.65 eV for interlayer hops \cite{ShinCM2016}.
They further found that if Li$^+$ migrates adjacent to a Li divacancy, the intralayer
barrier drops to about 0.38 eV and the interlayer barrier to about 0.18 eV, while in the
presence of a Li trivacancy the intralayer barrier can fall to $\sim 0.05$ eV and the
interlayer hop becomes effectively barrierless \cite{ShinCM2016}.
When compared with the microscopic (0.156 eV) and macroscopic (0.46 eV) experimental activation energies 
discussed above \cite{Sugiyama2013,Nakamura20101359}, the barriers for Li+ migration 
in the presence of a single vacancy warrant further examination.
Moreover, Li divacancies and trivacancies are unlikely to be abundant in the
low-delithiation limit of nearly stoichiometric powder samples, such as those employed in
the $\mu^+$SR and ac-impedance measurements of Sugiyama \emph{et al.} and Nakamura
\emph{et al.} \cite{Sugiyama2013,Nakamura20101359}.

A second limitation of much of the existing first-principles literature on Li$_2$MnO$_3$
is its treatment of magnetism.
In previous DFT+$U$ studies, calculations have been performed within spin-polarized
DFT+$U$ assuming ferromagnetic order
\cite{XiaoCM2012,HungruCM2016,WangJMCA2017}, in magnetically ordered states with the
underlying spin configuration left unspecified \cite{ShinCM2016,Sarkar2017}, or in a
fully nonmagnetic (spin-unpolarized) state \cite{HikimaJACS2022,SeoNchem2016}.
However, Li$_2$MnO$_3$ has a relatively low N\'eel temperature, $T_N \approx 36$~K
\cite{StrobelJSSC1988,LeeJPCM2012}, and is paramagnetic under the conditions relevant
for Li-ion diffusion experiments and battery operation.
Nonmagnetic calculations suppress local Mn moments entirely, whereas a true paramagnet
exhibits zero net magnetization only as a result of thermal fluctuations of finite local
moments.
A finite-temperature framework that preserves local-moment physics- such as DFT combined
with dynamical mean-field theory (DFT+DMFT)
\cite{ATLeePRB2023,MOLENDA200273,KorotinPRB2019,IsaacsPRB2020} -  is therefore required
to describe diffusion barriers in the experimentally relevant paramagnetic phase.

Owing to the low crystallographic symmetry of Li$_2$MnO$_3$ (space group $C2/m$, No.~12), 
standard Wannier projections in the DFT+DMFT framework often produce effective Hamiltonians 
with substantial off-diagonal matrix elements, which can hinder DMFT from correctly reproducing the insulating gap of Li$_2$MnO$_3$.
In our recent DFT+DMFT study of Li$_2$MnO$_3$ \cite{ATLeePRB2023}, we identified an efficient strategy to accurately capture 
its electronic structure by diagonalizing the Mn-$d$ block. 
This approach demonstrates the viability of a low-energy, $d$-only model for Li$_2$MnO$_3$, 
in which the resulting Wannier basis implicitly encodes Mn-$d$--O-$p$ hybridization.

In this work, we address these issues by combining DFT+$U$, DMFT with a quantum Monte Carlo
impurity solver, and NEB calculations for a single Li vacancy in paramagnetic Li$_2$MnO$_3$
at room temperature.
We first construct migration pathways using DFT+$U$ NEB for six representative intra- and
interlayer hops, and then evaluate DMFT total energies along these geometries, allowing the
correlated electronic configuration to relax at each image.
Dynamical electronic correlations substantially reduce the activation energies for the
lowest-barrier paths, yielding $E_a = 0.18$ eV for the lowest-$E_a$ intralayer hop and
$E_a = 0.50$ eV for the next-lowest intralayer path, while the barriers for the remaining
paths within DMFT remain close to their DFT+$U$ values.
These two lowest barriers quantitatively reproduce the short-range activation energy from
$\mu^+$SR and the long-range activation energy from ac-impedance measurements, establishing
a unified microscopic picture without invoking clustered vacancy configurations or strong
extrinsic disorder.

\begin{figure}
\begin{center}
\includegraphics[width=0.4\textwidth, angle=0]{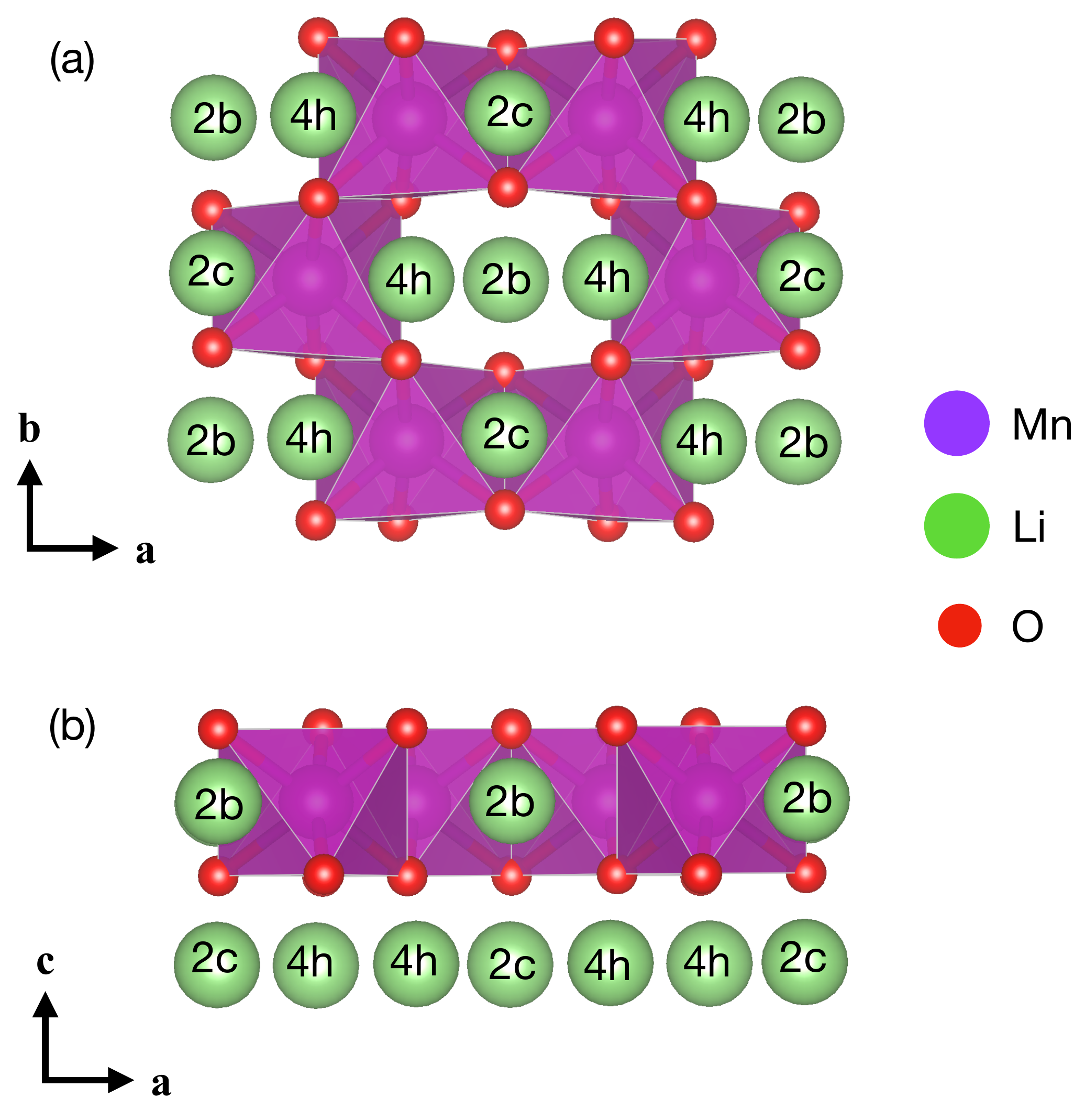}
\caption{ Atomic structure of Li$_2$MnO$_3$ with positions of Li ions.}
\label{atm_str}
\end{center}
\end{figure}

\section{Computational methods}

\subsection{DFT+DMFT}
\label{sec:method_dmft}

We employ the non-charge-self-consistent DFT+DMFT method
\cite{PRB2020Hyowon,DMFTwDFT} for DFT-relaxed structures.
For the underlying DFT calculations, we use the projector augmented wave
(PAW) method \cite{PAW} and the revised generalized gradient
approximation (GGA) functional proposed by Perdew {\it et al.}
(PBEsol) \cite{PBEsol}, as implemented in VASP \cite{VASP}. A
spin-independent version of the exchange-correlation functional is
employed. A plane-wave basis with a kinetic-energy cutoff of 500 eV is
used. We used 48-atom supercells (i.e., $1\times 2\times 2$ unit cells),
which contain 8 Mn atoms, together with $\Gamma$-centered
\textbf{k}-point meshes of size $10\times5\times5$. Atomic positions
were relaxed until the residual forces were less than 0.01 eV/\AA, and
the residual stress was reduced below 0.02 kB.

We solve the many-body problem in the Mn $3d$-only manifold. The
DFT+DMFT calculation proceeds as follows. First, we solve the
non-spin-polarized Kohn--Sham equations within DFT using VASP. Second,
we construct a localized-basis Hamiltonian for the Mn $3d$ bands by
generating maximally localized Wannier functions (MLWFs)\cite{MLWF} from
the nonmagnetic DFT band structure. For the $d$-only basis, the energy
window is chosen from $E_F-2.0$ eV to $E_F+4.0$ eV. For the main
DFT+DMFT calculations, we use $U=2.5$ eV and $J=0.9$ eV for the Mn-$d$
correlated subspace. This choice is consistent with our previous
$d$-only study, which showed that $U=2.0$--2.5 eV reproduces the
experimental band gap $E_g$.\cite{ATLeePRB2023} Finally, we solve the
DMFT self-consistent equations for the correlated subspace of Mn $3d$
Wannier orbitals using the continuous-time quantum Monte Carlo (CTQMC)
impurity solver.\cite{PRBHaule2007,ctqmcRMP2011}

In the present non-charge-self-consistent DFT+DMFT implementation,
total energies are available but ionic forces are not. Therefore, the
migration pathways and saddle-point geometries used in this work were
optimized at the DFT+$U$ level. The DFT+DMFT total energies were then
evaluated on the corresponding fixed NEB geometries for the initial,
saddle-point, and final images, after convergence of the correlated
electronic state for each configuration. Accordingly, the present
workflow is designed to assess how finite-temperature dynamical
correlations renormalize the energetics along a common ionic pathway,
rather than to perform a fully DFT+DMFT-relaxed NEB optimization.


For the initial, saddle-point, and final images, we tested several different initial conditions for the DFT+DMFT calculation, 
differing in the choice of Mn-centered correlated subspace used to construct the initial guess, 
because the $d$-only DFT+DMFT setup can converge to multiple low-energy solutions for the same ionic geometry.
We retained the lowest-energy converged solution
for the subsequent analysis. In practice, the numerical convergence of
the CTQMC-based DFT+DMFT calculations was monitored from the stability
of the impurity-energy contribution to the total energy over successive
DMFT iterations. Once its residual oscillation was within approximately
10 meV, the solution was regarded as converged, and the final total
energy was obtained by averaging over the last 10 DMFT iterations. The
impurity occupancy $N_d$ was typically even more stable, varying only at
the level of $\sim 10^{-3}$.

The rotationally invariant Coulomb interaction in the form of the Slater-Kanamori interaction Hamiltonian \cite{Slater-1951, Kanamori-1963, Kanamori-overview} is
\begin{equation}
\begin{split}
\hat{H}_\textrm{SK} =  \ &U \sum_{\alpha} \hat{n}_{\alpha\uparrow} \hat{n}_{\alpha\downarrow}
+\frac{1}{2}\sum_{\alpha \neq \beta} \sum_{\sigma \sigma'} \left( U' - J \delta_{\sigma \sigma'} \right) 
\hat{n}_{\alpha\sigma} \hat{n}_{\beta\sigma'} \\
 &-  \sum_{\alpha \neq \beta} \left( 
J c^{\dagger}_{\alpha\uparrow} c_{\alpha\downarrow} c^{\dagger}_{\beta\downarrow} c_{\beta\uparrow} +
J' c^{\dagger}_{\beta\uparrow} c^{\dagger}_{\beta\downarrow} c_{\alpha\uparrow} c_{\alpha\downarrow} 
\right). 
\end{split}
\label{sk-ham}
\end{equation}
Here, $c_{\sigma}$ and $c^{\dagger}_{\sigma}$ denote the fermion annihilation 
and creation operators, where $\sigma$ is the spin.
$U$ denotes an intra-orbital density-density interaction parameter, 
$U'$ is an inter-orbital density-density interaction parameter,
$J$ is a spin-flip interaction parameter, and 
$J'$ is a pair-hopping interaction parameter.
$U'=U-2J$ and $J'=J$ are due to  rotational invariance.


Within DFT+DMFT framework \cite{DMFTwDFT}, the self-energy convergence is achieved when 
$\Sigma^{\textrm{loc}} ( i \omega) = \Sigma^{\textrm{imp}} (i \omega)$, where 
$\Sigma^{\textrm{loc}} ( i \omega)$ and $\Sigma^{\textrm{imp}} (i \omega)$
are local and lattice self-energies, respectively, and $i\omega$ is imaginary frequency.
$\Sigma$ is approximated as a local quantity in the correlated subspace.
DFT+DMFT total energy is given by 
\begin{equation}
 \label{eq:E_tot}
\begin{split}
    E^{\textrm{TOT}} =& E^{\textrm{DFT}} (\rho) 
    + \sum_{m, \mathbf{k}} \epsilon_{m}(\mathbf{k}) \cdot \big [ n_{mm}(\mathbf{k}) - f_{m}(\mathbf{k}) \big ] \\
    &+ E^{\textrm{POT}} - E^{\textrm{DC}},
\end{split}
\end{equation}
where $E^{\textrm{DFT}}$ is the total energy from non spin-polarized DFT, and 
$\epsilon_{m}(\mathbf{k})$ are the DFT eigenvalues.
$n_{mm}(\mathbf{k})$ and $f_{m}(\mathbf{k})$ are the diagonal DMFT occupancy matrix element 
and Fermi function, respectively, for $m$th KS band and momentum $\mathbf{k}$.
The potential energy $E^{\textrm{POT}}$ is calculated by using Migdal-Galiski formula \cite{Galitskii}: 
\begin{equation}
    E^{\textrm{POT}} = \frac{1}{2}  \sum_{\omega} \big [ \Sigma^{\textrm{loc}}(i\omega) \cdot G^{\textrm{loc}} (i\omega)].
\end{equation}
Here, the local Green's function is simplified by 
$G^{\textrm{loc}} (i\omega) = \sum_{\mathbf{k}} G^{\textrm{loc}} (\mathbf{k},i\omega)$.

To obtain the spectral function, the maximum entropy method is used for the analytic continuation.
Spectral function $A(\omega)$ is given by
\begin{equation}
    A(\omega) = -\frac{1}{\pi} \textrm{Im} \Big [ \sum_{\mathbf{k}} G^{\textrm{loc}}(\mathbf{k}, \omega)\Big].
\end{equation}
%

Large off-diagonal terms in the Hamiltonian can lead to significant errors 
within the DMFT method, as continuous time quantum Monte Carlo (CTQMC) only treats the diagonal terms 
to circumvent the sign problem. 
The non-parallel alignment of the cartesian axes of the Wannier orbitals 
and the directions of the Mn-O bonds arises from the $C_{2h}$ point group symmetry 
of the MnO$6$ octahedron [Fig. \ref{atm_str}]. 
In cases where the point group symmetry of the transition metal (TM) octahedron 
is non-cubic, such as trigonal or monoclinic, there is a substantial mixing of the 
$d$ basis ($d_{xy}$, $d_{xz}$, $d_{yz}$, $d_{z^2}$, $d_{x^2-y^2}$), 
as these bases are defined within the cubic crystal field framework. 
Consequently, the off-diagonal terms of the Wannier Hamiltonian with the cubic 
$d$ orbital basis become significant, leading to errors within the DMFT calculations. 
To resolve this issue, we diagonalize Mn $d$ blocks of the Hamiltonian by applying a unitary rotation matrix,
which is successful to reproduce the experimental energy gap  \cite{ATLeePRB2023}.

Within DMFT, number of $d$ electrons in Mn ($N_d$), is computed from the local Green function 
$G^{\textrm{loc}}(\mathbf{k}, \mathbf{k}^{\prime},i \omega)$:
\begin{equation}
\label{Nd1}
    N_d = \sum_{a, \omega} \sum_{ \mathbf{k}, \mathbf{k}^{\prime}}  \textrm{Im} \big \{  [ \phi^a_d(\mathbf{k})]^{\ast} G^{\textrm{loc}}(\mathbf{k}, \mathbf{k}^{\prime}, i \omega) \phi^{a}_d (\mathbf{k}^{\prime}) \big \},
\end{equation}
where $\phi^a_d(\mathbf{k})$ is the normalized $d$-orbital wavefunction, 
which is transformed from the wavefunction in the real space
$\phi^a_d(\mathbf{r})$ with the center of coordinates 
$\mathbf{r}$ on a transition metal ion.

\subsection{DFT+$U$ and atomic structures}
\label{sec:method_dft}

The GGA+$U$ scheme within the rotationally invariant formalism together with the fully localized
limit double-counting formula \cite{LDA+U1} is used to study the effect of electron interactions.  
We used $U$=4 and $J$= 0 unless specified.
Projected density of states (PDOS) are obtained by the spherical harmonic projections 
inside spheres around each atom.
Wigner-Seitz radii of 1.323 \AA~ were used for the projection of Mn atoms, 
respectively, as implemented in the VASP-PAW pseudopotential.

Both spin-unpolarized and spin-polarized versions of the exchange–correlation functional 
were employed in the DFT+$U$ calculations. 
For each magnetic configuration, the structure was fully relaxed, including both 
internal atomic coordinates and lattice stresses. 
We used a 1$\times$1$\times$2 supercell to compare the relative energies of different 
magnetic phases within DFT+$U$, and a 1$\times$2$\times$2 supercell to compute 
Li migration paths in both the DFT+$U$ and DMFT calculations.

\section{Results and discussion}

\subsection{Electronic structure of Li$_2$MnO$_3$ with single Li vacancy}

\begin{figure}[t]
\begin{center}
\includegraphics[width=0.48\textwidth, angle=0]{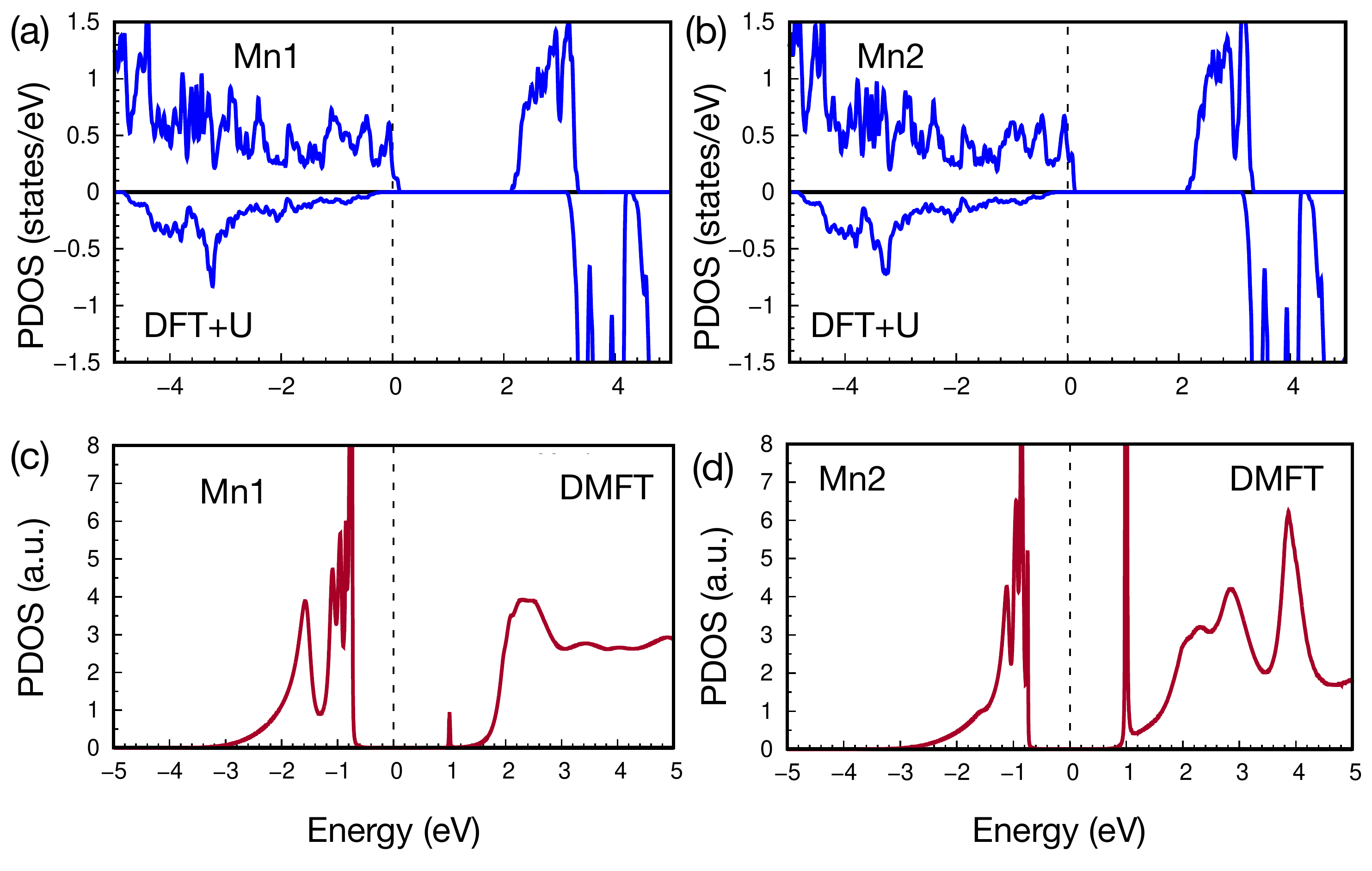}
\caption{Electronic structure of Li$_2$MnO$_3$ with a single Li vacancy from DFT+$U$ and
DMFT. (a,b) DFT+$U$ projected density of states (PDOS) onto Mn1 and Mn2 $d$ states.
(c,d) DMFT PDOS projected onto the Mn-$d$--like correlated subspace on Mn1 and Mn2.
Mn1 and Mn2 denote inequivalent Mn sites.
We caution that in the downfolded DMFT setup the correlated orbitals are Mn-$d$--like Wannier
functions that implicitly include Mn--O hybridization; therefore changes in the projected
Mn-$d$ occupancy reflect the effective low-energy description and should not be interpreted
as a direct oxidation-state assignment in the full $p$--$d$ manifold.}

\label{fig:PDOS}
\end{center}
\end{figure}

First, we study the electronic properties of Li$_2$MnO$_3$ with a single Li vacancy using DFT+$U$. 
There are three distinct Li sites, 2$b$, 4$h$, and 2$c$, as shown in Fig.~\ref{atm_str}.
Using a $1\times1\times2$ supercell containing four Mn atoms, we compare the total energies of 
four collinear magnetic configurations: ferromagnetic (FM), A-type antiferromagnetic (A-AF), 
C-type antiferromagnetic (C-AF), and G-type antiferromagnetic (G-AF), for different Li-vacancy 
positions, where A-, C-, and G-type denote antiferromagnetic alignment along one, two, and three
crystallographic directions defined by the conventional $a$, $b$, and $c$ lattice vectors.
For each Li-vacancy configuration, the FM phase is more stable than the AF phases by 14--182 meV
(see Table I in Appendix).
Thus, in the Li-diffusion calculations employing a larger $1\times2\times2$ supercell, we adopt the FM configuration.

Figure~\ref{fig:PDOS} shows the projected density of states (PDOS) onto Mn $d$ states in
Li$_2$MnO$_3$ with a single Li vacancy.
Removing one Li atom reduces the electron count by one (creating a single hole).
We examine two inequivalent Mn sites: Mn1, adjacent to the vacancy
(Mn1--Li$_{\mathrm{vac}}=2.89$~\AA), and Mn2, farther away
(Mn2--Li$_{\mathrm{vac}}=3.90$~\AA).
Within DFT+$U$, the Mn-projected spectra at Mn1 and Mn2 are nearly indistinguishable
[Fig.~\ref{fig:PDOS}(a,b)]: no sharp in-gap state forms, and the weight near the
valence-band maximum (VBM) remains broadly distributed.
Consistently, the integrated Mn-$d$ occupations vary only weakly,
$N_d(\mathrm{Mn1})=4.88$ and $N_d(\mathrm{Mn2})=4.89$, indicating that the vacancy does not
produce a strongly site-selective change within the Mn-$d$ manifold at this level of theory.
In line with prior DFT+$U$ and hybrid-functional studies
\cite{XiaoCM2012,SeoNchem2016,PRA2015Hoang},
the vacancy-induced change at the VBM is primarily associated with O~$2p$ states (ligand-hole
character), rather than the formation of a well-defined Mn$^{5+}$ oxidation state.

In contrast, the $d$-only DMFT results [Fig.~\ref{fig:PDOS}(c,d)] show a strongly
site-dependent redistribution of spectral weight within the Mn-$d$--like correlated
subspace.
Relative to Mn1, Mn2 exhibits a pronounced depletion of occupied Mn-$d$ weight together
with an increase of unoccupied weight near $E_F$, while the Mn1-projected spectrum remains
comparatively unchanged.
The corresponding Mn-$d$--like occupations are $N_d(\mathrm{Mn1})=3.00$ and
$N_d(\mathrm{Mn2})=2.01$ (nominal values within the reduced correlated subspace), i.e.,
the dominant reduction of the Mn-$d$--like occupation occurs on Mn2.
We emphasize that in this downfolded description the O degrees of freedom are not treated
explicitly; their influence is incorporated through renormalized crystal fields and
hoppings of the Mn-$d$--like Wannier basis.
Therefore, the apparent Mn-centered site dependence should be interpreted as a property of
the effective low-energy representation, not as direct evidence for a literal
integer-valence Mn oxidation-state change in the full $p$--$d$ electronic structure.
A definitive real-space assignment of oxygen- versus manganese-centered hole character would
require a fully charge-self-consistent $p$--$d$ correlated subspace, which is considerably
more computationally demanding and is left for future work.
This difference between DFT+$U$ and DMFT in the correlated-subspace charge rearrangement in
the presence of the vacancy also suggests that the two methods can yield different Li-ion
migration barriers.
Since DMFT modifies the electronic response along the migration coordinate via dynamical
correlations (as quantified below through the total-energy decomposition), the resulting
migration barrier can differ from that obtained in static DFT+$U$.

\subsection{Li-ion migration barriers: DFT+$U$ versus DMFT}

\begin{figure}[!t]
\begin{center}
\includegraphics[width=0.47\textwidth, angle=0]{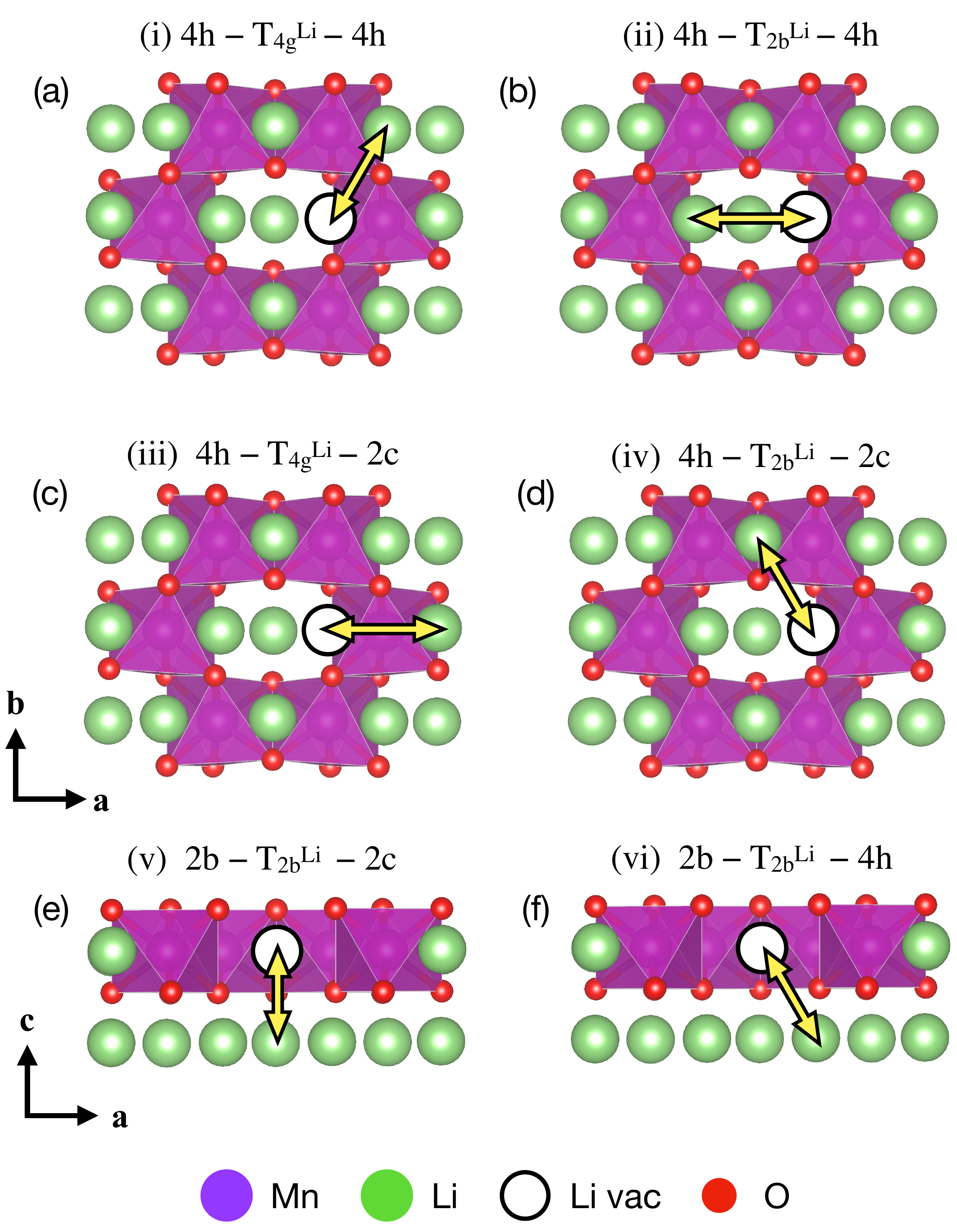}
\caption{ Diffusion path of single Li vacancy in Li$_2$MnO$_3$.
(a)-(d) intra layer diffusion. (e)-(f) inter layer diffusion.
}
\label{neb_path}
\end{center}
\end{figure}

\begin{figure}[!t]
\begin{center}
\includegraphics[width=0.40\textwidth, angle=0]{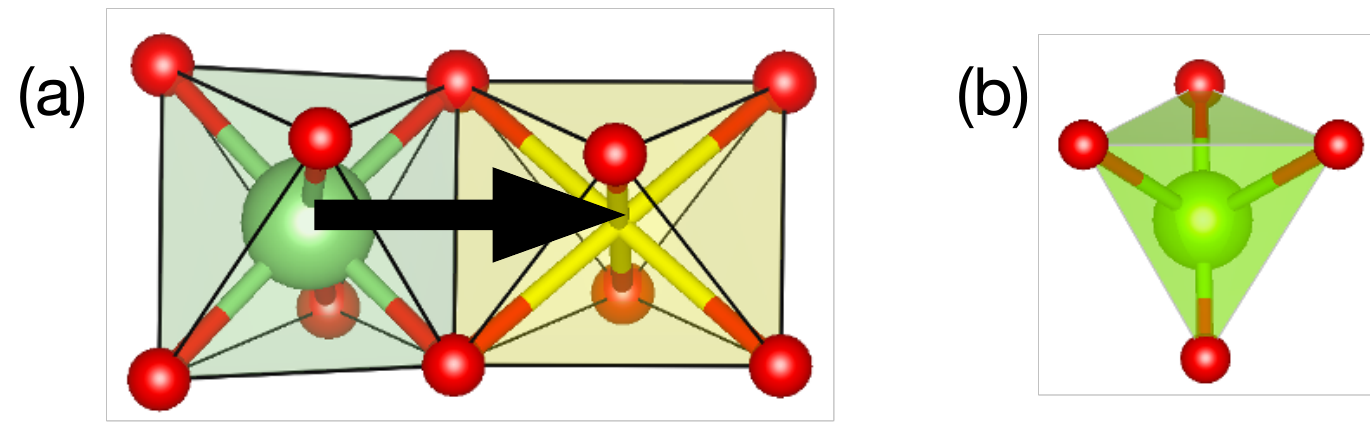}
\caption{ (a) Diffusion path (\emph{ii}) of single Li vacancy in Li$_2$MnO$_3$,
and (b) LiO$_4$ tetrahedral structure at the saddle point.
}
\label{saddle_config}
\end{center}
\end{figure}

Next, we analyze Li$^+$ migration in Li$_2$MnO$_3$ using DFT+$U$ and
DFT+DMFT. We consider six symmetry-inequivalent pathways defined in
Fig.~\ref{neb_path}: four \emph{intralayer} hops,
(i) $4h - T^{\mathrm{Li}}_{4g} - 4h$,
(ii) $4h - T^{\mathrm{Li}}_{2b} - 4h$,
(iii) $4h - T^{\mathrm{Li}}_{4g} - 2c$,
(iv) $4h - T^{\mathrm{Li}}_{2b} - 2c$,
and two \emph{interlayer} hops,
(v) $2b - T^{\mathrm{Li}}_{2b} - 2c$,
and (vi) $2b - T^{\mathrm{Li}}_{2b} - 4h$.
These paths are similar to those considered in a previous DFT+$U$
study.\cite{ShinCM2016}

\begin{figure}[!t]
\begin{center}
\includegraphics[width=0.48\textwidth, angle=0]{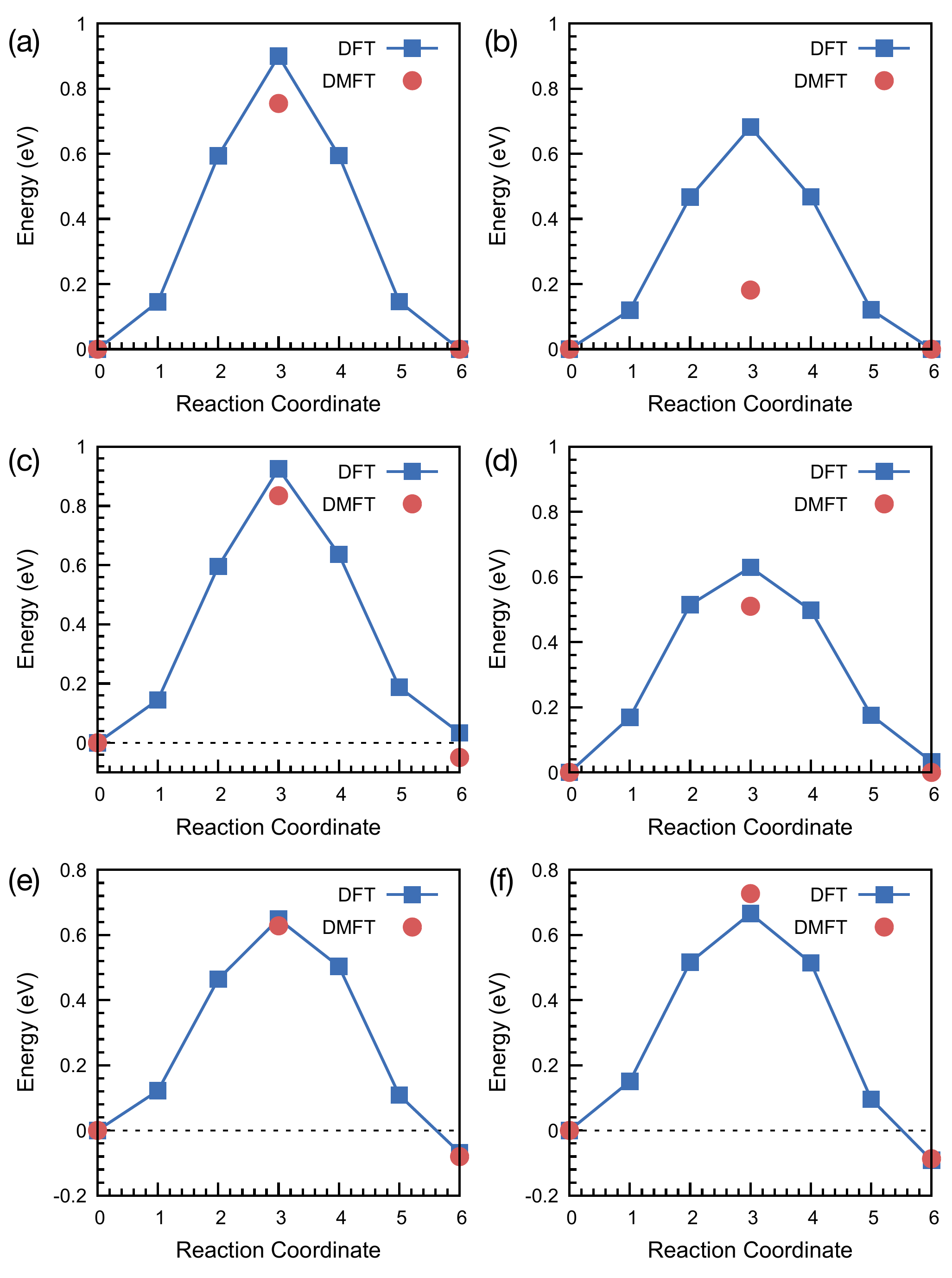}
\caption{ (a)–(f) Migration barriers for a single Li vacancy in Li$_2$MnO$_3$ along pathways (i)–(vi), respectively. 
}
\label{Ea}
\end{center}
\end{figure}

The corresponding activation energies $E_{\mathrm a}$ are summarized in
Fig.~\ref{Ea} (blue: DFT+$U$; red: DFT+DMFT). We first obtain the
migration pathways and saddle-point geometries using NEB within
spin-polarized DFT+$U$, and then evaluate the DFT+DMFT total energies of
the initial, final, and saddle-point configurations on these fixed
geometries. Because the present DFT+DMFT implementation does not provide
ionic forces, the present comparison does not constitute a fully
DFT+DMFT-relaxed NEB calculation. Instead, it isolates how
finite-temperature dynamical correlations modify the relative energetics
along a common migration pathway. For each NEB image, we also tested
multiple initial guesses and retained the lowest-energy converged
DFT+DMFT solution in the subsequent analysis.

Within DFT+$U$, the migration barriers for intralayer diffusion are $E_a = 0.63$–0.93 eV.
In all cases, Li$^+$ migrates via the tetrahedral interstitial site ($T$), forming a LiO$_4$
tetrahedron at the saddle point, as shown in Fig.~\ref{saddle_config}, consistent with
previous work \cite{ShinCM2016}.
The barriers along paths (i) and (iii) are relatively large, $E_a = 0.90$ and 0.93 eV,
respectively, in reasonable agreement with the earlier DFT+$U$ values of 0.78 and 0.80 eV
\cite{ShinCM2016}; the remaining differences are likely due to details of the computational
setup (e.g., supercell size and choice of parameters).
In contrast, the barriers along paths (ii) and (iv) are lower, $E_a = 0.68$ and 0.63 eV,
indicating that path (iv) provides the lowest intralayer migration barrier.
The earlier DFT+$U$ study also found $E_a = 0.60$ and 0.56 eV for paths (ii) and (iv),
respectively, likewise identifying path (iv) as the lowest-barrier pathway.
For interlayer diffusion within DFT+$U$, the barriers are $E_a = 0.65$ and 0.67 eV for
paths (v) and (vi), respectively, which are slightly larger than the lowest intralayer
diffusion barrier.

Next, we evaluate $E_a$ within DFT+DMFT at 300 K in the paramagnetic
phase. As in DFT+$U$, the lowest-$E_a$ diffusion paths are (ii)
$4h-T_{2b}^{\mathrm{Li}}-4h$ and (iv) $4h-T_{2b}^{\mathrm{Li}}-2c$, but
DFT+DMFT notably reduces the corresponding barriers, as shown in
Fig.~\ref{Ea}. The DFT+DMFT activation energies for paths (ii) and (iv)
are 0.18 and 0.50 eV, respectively, while for the remaining paths the
DFT+$U$ and DFT+DMFT barriers differ by only 0.02--0.15 eV. We note
that, in the present DFT+DMFT setup, the correlated subspace is defined
by Mn-$d$-like Wannier functions that implicitly include Mn--O
hybridization; consequently, the vacancy-induced charge rearrangement is
described within this effective low-energy manifold.

As a limited check of the interaction-parameter dependence, we also
performed additional DFT+DMFT calculations for a smaller
$1\times1\times2$ supercell by varying the Mn-$d$ interaction strength
from $U=2.0$ to $U=2.5$ eV at fixed $J=0.9$ eV. We focus on this range
because, in our previous $d$-only study, $U=2.0$--2.5 eV reproduced the
experimental band gap reasonably well.\cite{ATLeePRB2023} For
representative migration pathways in the smaller supercell, the resulting
activation energies change only modestly over this range of $U$,
typically by at most a few hundredths of an eV. Thus, within this
physically motivated parameter range, increasing $U$ leads only to a
modest quantitative change in $E_{\mathrm{a}}$. A full
interaction-parameter scan for the larger supercell used in the main
text was not attempted because of the computational cost.

The lowest barrier of 0.18~eV corresponds to a short-range Li jump
between neighboring sites within the Li$_2$MnO$_3$ region. This value is
close to the experimental activation energy $E_a = 0.156$~eV obtained
from $\mu^+$SR measurements, which probe local Li self-diffusion in
nearly stoichiometric Li$_2$MnO$_3$ powders~\cite{Sugiyama2013}. Our
DFT+DMFT results therefore suggest that the experimentally observed
short-range activation scale can be rationalized by single-vacancy
migration along the lowest-barrier pathway in the paramagnetic phase,
without invoking Li divacancies or trivacancies.

From ac-impedance measurements on Li$_2$MnO$_3$ powder samples, which
probe long-range Li-ion transport, Nakamura \emph{et al.} reported an
activation energy of $E_a = 0.46$~eV at 900 \unit{\degreeCelsius}
\cite{Nakamura20101359}. In our DFT+DMFT barrier landscape, the
0.18~eV hop alone does not form a percolating pathway for net Li
transport out of the Li$_2$MnO$_3$ region [Fig.~\ref{Ea}(b)], so
long-range migration requires additional hops. Among these, the
intralayer path (iv) has the lowest barrier, $E_a = 0.50$~eV, whereas
the other relevant paths are higher. Thus, an energetically favorable
long-range route consists of many low-barrier 0.18~eV hops combined
with occasional 0.50~eV hops, for example
$4h - T^{\mathrm{Li}}_{2b} - 4h - T^{\mathrm{Li}}_{2b} - 2c$, and the
overall rate is controlled by the larger barrier, $E_a \approx
0.50$~eV. Despite the different experimental temperature, this
controlling barrier is of the same order as the activation energy
extracted from ac-impedance measurements, indicating a comparable
bottleneck scale for long-range transport.


To clarify the origin of the reduced $E_a$ within DFT+DMFT, we analyze
the change in the total energy $E^{\mathrm{TOT}}$ [Eq.~(\ref{eq:E_tot})]
between the initial and saddle-point configurations for the
lowest-barrier path, path~(ii). For the underlying DFT contribution
$E^{\mathrm{DFT}}$, the saddle point lies 0.678~eV above the initial
state. By contrast, the DFT+DMFT eigenvalue-correction term,
$\sum_{m,\mathbf{k}} \epsilon_m(\mathbf{k})
\left[n_{mm}(\mathbf{k})-f_m(\mathbf{k})\right]$,
substantially stabilizes the saddle point: this term is lower at the
saddle point than at the initial configuration by 0.506~eV, thereby
providing the dominant reduction of the net migration barrier. The
remaining terms in Eq.~(\ref{eq:E_tot}) are comparatively small:
$E^{\mathrm{POT}}$ is higher at the saddle point by only 0.070~eV, and
$E^{\mathrm{DC}}$ changes by merely $-0.45$~meV. Thus, for path~(ii),
the barrier lowering within DFT+DMFT is dominated by the
eigenvalue-correction term rather than by changes in the interaction or
double-counting energies.

A useful contrast is provided by path~(iii), for which the underlying
DFT contribution to the barrier is similarly large, 0.790~eV, but the
DFT+DMFT corrections remain very small. In this case, the
eigenvalue-correction term changes by only $+0.028$~eV between the
initial and saddle-point configurations, while $E^{\mathrm{POT}}$ and
$E^{\mathrm{DC}}$ vary by just 0.006 and 0.004~eV, respectively. The
comparison between paths~(ii) and~(iii) therefore shows that the
pronounced suppression of $E_a$ is strongly path dependent. In
particular, path~(ii) exhibits a large negative eigenvalue correction,
whereas path~(iii) does not, despite their similarly large underlying
DFT barriers.

This difference correlates with the local registry of the migrating Li
at the saddle point relative to the surrounding MnO$_6$ network. For
path~(ii), the migrating Li passes through a relatively open region of
the Li sublattice, near a Li-centered hexagonal arrangement, so that
there is no Mn ion located directly above or below the saddle-point Li.
Consistent with this geometry, the nearest Li--Mn distance is relatively
long, 3.254~\AA. By contrast, for path~(iii) the saddle-point Li lies
much more directly beneath nearby MnO$_6$ octahedra, with two short
Li--Mn distances of 2.514 and 2.596~\AA. Path~(iv) appears intermediate
between these limits: its saddle-point Li also passes near a
Li-centered hexagonal region, but with a less open local environment
than in path~(ii), and its shortest Li--Mn distance is correspondingly
intermediate at 3.030~\AA. 
Thus, the local registry of the saddle-point Li relative to the MnO$_6$ network correlates 
with the magnitude of the DFT+DMFT barrier renormalization, suggesting that the 
more open local environment of path~(ii) is more effectively stabilized within the correlated electronic 
description than the corresponding saddle points of paths~(iii) and~(iv).


\section{Conclusion}

In this work, we investigated Li-ion migration in Li$_2$MnO$_3$ by combining DFT+$U$,
DFT+DMFT (with a continuous-time quantum Monte Carlo solver), and nudged-elastic-band (NEB)
calculations for a single Li vacancy in the paramagnetic phase. By evaluating DMFT total
energies on the fixed DFT+$U$ NEB geometries for six representative intra- and interlayer
paths---and optimizing the correlated electronic state at each NEB image---we find that
dynamical correlations reduce the activation energies for the lowest-barrier paths to
$E_a = 0.18$~eV for path (ii) and $E_a = 0.50$~eV for path (iv), while leaving the other
paths largely unchanged relative to DFT+$U$. The lowest barrier, $E_a = 0.18$~eV, is close to
the microscopic activation energy for short-range Li self-diffusion extracted from $\mu^+$SR
measurements on nearly stoichiometric Li$_2$MnO$_3$ powders, suggesting that the local
activation scale can be rationalized by single-vacancy migration along path (ii) in a
paramagnetic host, without invoking Li divacancies or trivacancies.

Long-range transport, on the other hand, necessarily involves sequences of hops that combine
low-barrier (0.18~eV) and higher-barrier (0.50~eV) steps; within such a network, the effective
activation energy is controlled by the larger barrier, $E_a \approx 0.50$~eV. Despite the
different experimental temperature, this controlling scale is of the same order as the
macroscopic activation energies obtained from ac-impedance measurements on ceramic powders.
Our results also indicate that interpretations based solely on interlayer diffusion are too
restrictive, and that a mixture of intralayer processes is required to reconcile
local and macroscopic probes.

These findings highlight the importance of beyond-DFT treatments for ionic diffusion in
correlated oxides, and motivate future extensions to higher vacancy concentrations and
mixed-cation Li-rich layered cathodes. In particular, extending the present framework to
regimes where oxygen redox becomes active will be essential for assessing how lattice defects,
charge redistribution, and dynamical correlations jointly influence transport and stability.

\section{Acknowledgment}
We acknowledge financial support from the U.S. Department of Energy, Office of Science, Office of Basic Energy Sciences, Materials Science and Engineering Division.
We gratefully acknowledge the computing resources provided on Improv, a high-performance computing cluster operated by the Laboratory Computing Resource Center at the Argonne National Laboratory.

\bibliographystyle{apsrev4-2}
\bibliography{rsc}

\section{Appendix A}
\label{appendix:A}

\begin{table}[h]
  \centering
  \caption{Relative energies (in eV) of collinear magnetic configurations within DFT+$U$
  for different Li-vacancy positions in the $1\times1\times2$ supercell.
  Here $E$ denotes the total energy of a given configuration, 
  $\Delta E = E - E_{\mathrm{min}}$ is measured with respect to the global ground state
  (FM with a Li vacancy at the $2b$ site), and 
  $\delta E = E - E_{\mathrm{FM}}(\text{Li-vac})$ is measured with respect to the FM state
  for the same Li-vacancy position.}
  \label{table:mag_config}
  \begin{tabular}{cccc}
    \hline\hline
    Li-vac site & Magnetic order & $\Delta E$ (eV) & $\delta E$ (eV) \\
    \hline
    $2b$ & FM   & 0.000 & 0.000 \\
         & A-AF & 0.014 & 0.014 \\
         & C-AF & 0.140 & 0.140 \\
         & G-AF & 0.145 & 0.145 \\
             \hline
    $4h$ & FM   & 0.065 & 0.000 \\
         & A-AF & 0.079 & 0.065 \\
         & C-AF & 0.179 & 0.165 \\
         & G-AF & 0.196 & 0.182 \\
             \hline
    $2c$ & FM   & 0.043 & 0.000 \\
         & A-AF & 0.086 & 0.043 \\
         & C-AF & 0.218 & 0.175 \\
         & G-AF & 0.216 & 0.173 \\
    \hline\hline
  \end{tabular}
\end{table}

\end{document}